\documentclass[prd,aps,showpacs,twocolumn,superscriptaddress]{revtex4}

\usepackage{epsfig}
\usepackage{epsf}
\usepackage{bm}                 
\usepackage{amsmath}
\usepackage{amssymb}
\usepackage{dcolumn}
\usepackage{graphicx}
\usepackage{psfrag}

\begin{document}


\title{Relativistic quantum information processing with bosonic and
  fermionic interferometers}   

\author{Pieter Kok} \email{pieter.kok@hp.com}

\affiliation{Hewlett Packard Laboratories, Filton Road Stoke Gifford, 
Bristol BS34 8QZ, UK}

\author{Samuel L.\ Braunstein} 

\affiliation{Computer Science, University of York, York YO10 5DD, UK}

\date{\today}

\begin{abstract}\noindent
 We derive the relativistic transformation laws for the annihilation
 operators of the scalar field, the massive spin-1 vector field, the
 electromagnetic field and the spinor field. The technique developed here
 involves straightforward mathematical techniques based on fundamental
 quantum field theory, and is applicable to the study of entanglement
 in arbitrary coordinate transformations. In particular, it predicts
 particle creation for non-inertial motion. Furthermore, we present a
 unified description of relativistic transformations and
 multi-particle interferometry with bosons and fermions, which
 encompasses linear optical quantum computing. 
\end{abstract}


\maketitle

\noindent
In his last contribution to the quantum archive, called {\em Quantum
  Information and General Relativity} \cite{peres04}, Asher Peres
  wrote that ``when I was a young man, my thesis adviser was Nathan
  Rosen, and the subject was the existence of gravitational radiation
  in general relativity. Only much later, I seriously learnt quantum
  mechanics, and still much later information theory. I now want to
  return to my roots and try to combine all these subjects together.''
  It therefore seems appropriate to honour the memory of Asher Peres
  with a paper that describes aspects of quantum information theory
  for observers in a (general) relativistic setting.

\section{Introduction}

\noindent
As a fundamental physical theory, relativistic quantum information
theory (RQIT) has widespread applications, ranging from practical
tools for describing moving observers in quantum communication
protocols, to black hole thermodynamics. For example, it is hoped that
RQIT will play a central role in clock synchronization and (optical)
quantum communication between a ground station and a relativistically
moving satellite, as well as the resolution of the black hole
information paradox.

So far, work on RQIT has focused predominantly on relativistic
transformations of single-particle wave functions.  Early work by
Czachor defined the relativistic spin operator for a relativistic
description of the violation of Bell inequalities
\cite{czachor97}. Then Peres {\em et al}.\ showed that the reduced
density matrix for the spin of an electron is not a Lorentz invariant
scalar \cite{peres02}, indicating that there is a spin-momentum
interaction in Lorentz transformations. Subsequently, Alsing and Milburn 
determined the transformation properties of entangled particles in
momentum eigenstates \cite{alsing02}, and it was shown by Gingrich and
Adami that spin and polarization entanglement between two Gaussian
wave packets is transferred to momentum entanglement under Lorentz
boosts \cite{gingrich02,gingrich03}.

Many of these results have been obtained using Wigner's {\em little
group} formalism
\cite{alsing02,gingrich02,gingrich03,wigner39}. Since the
little group is constructed from a standard momentum four-vector that
is invariant under Lorentz transformations, this formalism breaks
down for {\em arbitrary} coordinate transformations. In addition, it
is not always straightforward to find  the representations of the
little group. For a wider application of relativistic results,
``plug-and-play'' transformation rules for the annihilation operator
seem more appropriate.

These transformation rules are essential for the description of
(multi-particle) quantum interferometry. In particular, we are
interested in a relativistic extension of linear optical quantum
computing. In this paper, we present a general technique for deriving
the annihilation operator of the scalar field, the spin-1 vector
field, the electromagnetic field, and the spin-$\frac{1}{2}$ Dirac
field for arbitrary coordinate transformations, and we give the
explicit result for Lorentz transformations. We then give a unified
description of the resulting Bogoliubov transformations and
multi-particle quantum interferometry. This will lead to a
relativistic formulation of linear optical quantum computing
\cite{knill01}.  

There are several advantages to our technique: {\em i}) The
transformation laws are relatively easy to obtain by integration,
rather than finding group representations. {\em ii}) The resulting
transformations are completely fundamental. {\em iii}) Our technique
is applicable to any coordinate transformation (including in the
presence of curvature), and is not restricted to Lorentz boosts. In
particular, it predicts particle creation for non-inertial
motion. {\em iv}) Substituting the transformed annihilation operator
into specific expressions of the state of a quantum field
automatically yields the correct transformed state. This way, it is
straightforward to describe boosted wave packets, and it allows us to
study the transformation of {\em entanglement in arbitrary coordinate
systems}. 

In the next section, we present a relativistic paradox, to put the
issue on edge. In section \ref{sec:ft}, we give the field
transformations for Lorentz-boosted quantum fields. In section
\ref{sec:int}, we solve the paradox, and sketch relativistic
multi-particle quantum interferometry and relativity. In principle,
this unification includes arbitrary particle creation associated with
non-inertial observers and observers on curved spacetime. Finally, we
present our conclusions in section \ref{sec:con}.

\section{the twin-photon paradox}


\noindent
In order to describe relativistic multi-particle quantum
interferometry, let's perform the following {\em gedanken experiment}:
Two single-photon plane waves with momenta $k_1$ and $k_2$, and
polarization $j\in\{ H,V\}$ meet at a 50:50 beam splitter. In the
outgoing modes of the beam splitter we place two ideal particle
detectors, which tell us with perfect fidelity how many photons there
are in that mode. It is well known that a 50:50 beam splitter causes
two identical photons to ``pair off'' into the outgoing modes. In
other words, we will never find any coincidence counts between the two
detectors. This is the so-called Hong-Ou-Mandel effect
\cite{hong87}. Mathematically, the transformation of the incoming
modes would be  
\begin{eqnarray}
 \hat{a}_j(k_1) &\rightarrow& \frac{1}{\sqrt{2}}\left[\hat{a}_j(k_1) +
 \hat{a}_j(k_2)\right] , \cr  
 \hat{a}_j(k_2) &\rightarrow& \frac{1}{\sqrt{2}} \left[-\hat{a}_j(k_1) +
 \hat{a}_j(k_2)\right] . 
\end{eqnarray}
The input state
$\hat{a}_j^{\dagger}(k_1) \hat{a}_j^{\dagger}(k_2) |0\rangle$ is then
transformed into $\frac{1}{2} [\hat{a}_j^{\dagger 2}(k_2) -
\hat{a}_j^{\dagger 2}(k_1)] |0\rangle$, where $|0\rangle$ denotes the
vacuum.

An observer in a boosted frame of reference will see quite a different
physical process taking place: To him, the two waves do not
necessarily have the same frequency and polarization. As a result, the
two photons are not identical, and there will be coincidence counts in
the detectors. However, photon counting yields numbers, which are
Lorentz invariant. Consequently, both observers should obtain exactly
the same detector statistics. Thus we arrive at a contradiction. 

A similar paradox can be constructed for fermions. Here, the exclusion
principle forbids identical particles from occupying the same quantum
state, resulting in the absence of two-fermion states in either output
mode of the fermionic beam splitter. Again, to a boosted observer the
fermions have different wavelengths and spin, leading to different
detector statistics. In order to resolve this paradox, we explicitly
calculate the transformation rules for the annihilation operator of
the quantum fields. Furthermore, both the twin-photon and the
twin-electron paradox is an example of two-particle quantum
interferometry. 

\section{Field transformations}\label{sec:ft}

\subsection{Scalar fields}

\noindent
The scalar quantum field $\phi(x)$ obeys the Klein-Gordon equation
$(\partial_{\mu} \partial^{\mu} - m^2) \phi = 0$, where $m$ is the
mass of the field, and Greek indices always denote components of a
four-vector. It can be expanded in terms of mode functions $f$:  
\begin{equation}\nonumber
 \phi(x) = \int \frac{d\,\mathbf{k}}{2k_0}\, \left[
   \hat{a}(k) f_k(x) + \hat{a}^{\dagger}(k) f^*_k(x) \right],
\end{equation}
 where $k$ is the four-momentum, $\mathbf{k}$ is the three-vector
 component of $k$, and 
$k_0$ is the energy component. The annihilation and creation operators
associated with mode $k$ are $\hat{a}(k)$ and $\hat{a}^{\dagger}(k)$
respectively. The annihilation operator is extracted using the 
time-independent inner product \cite{bjorken65}:
\begin{equation}
 \hat{a}(k) = i\int d^3 x\, f^*_k (x)\,
 \overset{\leftrightarrow}{\partial_0}\, \phi(x) \equiv (f_k,\phi)\; , 
\end{equation}
where $a(t)\, \overset{\leftrightarrow}{ \partial_0}\, b(t) = a(t)
\partial_0 b(t) - [\partial_0 a(t)] b(t)$. Typically, we choose the
plane-wave expansion $f_k (x) = [(2\pi)^3 2
  k_0]^{-\frac{1}{2}}e^{ikx}$, where $kx \equiv k_{\mu} x^{\mu}$.

Alice and Bob are two observers that occupy two different reference
frames. Alice describes the field $\phi(x)$ in terms of her
coordinates $x$, whereas Bob describes the field $\phi(x')$ in terms
of his coordinates $x'$. The two coordinate systems are connected by
an invertible transformation. In this paper we will restrict
ourselves to Lorentz transformations $\Lambda$ such that $x'= \Lambda 
(x-\ell)$, with $\ell$ an arbitrary translation. However, our results
also apply to arbitrary coordinate transformations corresponding to
non-inertial relative motion. In addition, Bob uses his own definition
of the annihilation and creation operators $\hat{a}' (k')$ and
$\hat{a}'^{\dagger} (k')$, and (plane wave) mode functions
$g(x')$. The question is now what are the transformation rules that
relate $\hat{a}(k)$ and $\hat{a}' (k')$. To  this end, we can extract
Bob's annihilation operator:   
\begin{equation}\label{tii}
 \hat{a}'(k') = i\int d^3 x'\, f^*_{k'} (x')\,
 \overset{\leftrightarrow}{\partial}_{0'}\, \phi(x')\; .
\end{equation}
Alternatively, Alice may describe the field $\phi$ in terms of her
coordinates $x(x')$. When we substitute this into Eq.~(\ref{tii})
and use $k' = \Lambda k$, we obtain  
\begin{equation}
 \hat{a}'(\Lambda k) = \hat{a} (k)\; e^{-i\Lambda k\ell}\; ,
\end{equation}
with a similar expression for the creation operators. It is clear that
the bosonic commutation relations still hold for 
$\hat{a}'$ and $\hat{a}'{}^{\dagger}$. The state of (multi-particle)
wave packets can be expressed in terms of a function $\mathsf{f}$ of creation
operators $\hat{a}^{\dagger}$ acting on a vacuum defined by $\hat{a}
|0\rangle = 0$. Similarly, the transformed annihilation operator
defines a vacuum state $\hat{a}' |0'\rangle = 0$. The state then
transforms as $\mathsf{f}(\hat{a}^{\dagger}) |0\rangle \rightarrow
\mathsf{f}(\mbox{$\hat{a}'$}^{\dagger}) |0'\rangle$. For Lorentz
transformations, the vacuum states of Alice and Bob are
identical. Other transformations, however, change the vacuum
state. Alice and Bob then no longer agree upon the number of particles
in the experiment.

In general, since quantum states can be expressed in terms of a
function of creation operators $\hat{a}^{\dagger}$ acting on the
vacuum $|0\rangle$, substituting this transformation rule
$\hat{a}'{}^{\dagger}$ on the vacuum $|0'\rangle$ will immediately yield
the correctly transformed quantum state.  

\subsection{Spin-1 massive boson fields}

\noindent
The simplest extension to the Klein-Gordon field is the spin-1 degree
of freedom, yielding a vector field $V^{\mu}(x)$ with mass $m$:
\begin{equation}\nonumber
 V^{\mu} (x) = \int \frac{d\,\mathbf{k}}{2k_0}\, \sum_{j=-1}^1 \left[
 \frac{\epsilon_j^{\mu} \hat{a}_j (k) e^{ikx}}{\sqrt{(2\pi)^3 2k_0}} +
 \mathrm{H.c.} \right] , 
\end{equation}
where $\epsilon^{\mu}_j$ is the four-vector associated with the
$j$-component of the field, and H.c.\ stands for Hermitian
conjugate. The field obeys the Lorentz gauge $\partial_{\mu} V^{\mu} =
k_{\mu} \epsilon_j^{\mu} = 0$, which, for a particle at rest suggests
the representation $k=(m,0,0,0)$, $\epsilon_1 = (0,1,0,0)$,
$\epsilon_0 = (0,0,1,0)$ and $\epsilon_{-1} = (0,0,0,1)$. The
relativistic transformation of the vector field is given by
\begin{equation}\nonumber
 {\Lambda^{\mu}}_{\nu} V^{\nu} (x) = \int \frac{d\,\mathbf{k}}{2k_0}\,
 \sum_j{\Lambda^{\mu}}_{\nu} \frac{\epsilon_j^{\nu}
 \hat{a}_j (k) e^{ikx}}{\sqrt{(2\pi)^3 2k_0}} + \mathrm{H.c}.
\end{equation} 
Extracting the annihilation operator using $f_k^{\mu}(x) =
[(2\pi)^3 2k_0]^{\frac{1}{2}}\epsilon_j^{\mu}\, e^{ikx}$ then yields 
\begin{equation}
 \hat{a}_j'(\Lambda k) = \sum_{l=-1}^1 \epsilon^*_{\mu,j}
     {\Lambda^{\mu}}_{\nu} \epsilon^{\nu}_l\; \hat{a}_l (k)\,
     e^{-i\Lambda   k\ell}.
\end{equation} 
Lorentz transformations do not leave three-volumes invariant, and we
need to renormalize the transformation to make it unitary. The boosted
annihilation operator then becomes 
\begin{equation}\label{spinone}
 \hat{a}_j'(\Lambda k) = -m^2 \sum_{l=-1}^1 \frac{\epsilon^*_{\mu,j}
 {\Lambda^{\mu}}_{\nu} \epsilon^{\nu}_l}{k_{\mu} {\Lambda^{\mu}}_{\nu}
 k^{\nu}}\; \hat{a}_l (k)\, e^{-i\Lambda k\ell} , 
\end{equation}
which obeys the bosonic commutation relations. Here, we observe a
boost-dependent change in spin. 

\subsection{Gauge fields}

\noindent
In order to find the proper Bogoliubov transformations for massless
spin-1 fields, it is clear from Eq.~(\ref{spinone}) that (contrary to
scalar fields) we cannot take the limit $m\rightarrow 0$. Massless
fields with spin, such as the quantized electromagnetic field, have an
extra gauge freedom that we need to take into account. Here, we
consider the vector potential of the electromagnetic field:
\begin{equation}\nonumber
 A^{\mu} (x) = \int \frac{d\,\mathbf{k}}{2k_0}\,
 \sum_j \left[ \frac{\epsilon_j^{\mu} \hat{a}_j (k)
 e^{ikx}}{\sqrt{(2\pi)^3 2k_0}} + \mathrm{H.c.} \right] ,
\end{equation}
where $j$ indicates two orthogonal polarizations. In addition to the
Lorentz gauge, it has to obey a second gauge relation, usually the
Coulomb gauge $\nabla \cdot \mathbf{A} = \mathbf{k} \cdot
\boldsymbol{\epsilon}_j = 0$. In this gauge, there is no longitudinal
polarization. Since $k_{\mu} \epsilon_j^{\mu}$ is an invariant scalar,
Lorentz transformations will keep the field in the Lorentz
gauge. However, this is not true for the Coulomb gauge, and since this
is the gauge that is typically used in the description of
multi-particle interferometry, we need to take this change into
account in our calculation.

The gauge freedom means that we can add the derivatives of two
massless Klein-Gordon scalar fields $\phi_j$ to the vector potential
in order to change the gauge:  
\begin{equation}\label{gauge}
 A^{\mu} \rightarrow A^{\mu} + \sum_j \alpha_j
 \partial^{\mu} \phi_j\; .
\end{equation}
The addition of such a gauge term does not change the observable outcomes,
since all physical observables depend only on derivatives of $A$, and
$\partial_{\mu} \partial^{\mu} \phi = 0$. 

The relativistic transformation of the vector potential is given by 
\begin{equation}\nonumber
 {\Lambda^{\mu}}_{\nu} A^{\nu} (x) = \int \frac{d\,\mathbf{k}}{2k_0}\,
 \sum_j {\Lambda^{\mu}}_{\nu} \frac{\epsilon_j^{\nu} 
 \hat{a}_j (k) e^{ikx}}{\sqrt{(2\pi)^3 2k_0}} + \mathrm{H.c}.
\end{equation}
By changing the coordinates $x = \Lambda^{-1} x'$ and changing the
integration variable $k = \Lambda^{-1} k'$, we find
\begin{equation}\nonumber
 {\Lambda^{\mu}}_{\nu} A^{\nu} (x') = \int
 \frac{d\,\mathbf{k}'}{2k_0}\, \sum_{\lambda}
 \Lambda_{\nu}^{\mu} \frac{\epsilon_j^{\nu} \hat{a}_j (\Lambda^{-1}
 k') e^{ik'x'}}{\sqrt{(2\pi)^3 2k_0'}} +  \mathrm{H.c}.
\end{equation} 
The Lorentz condition is still satisfied, as is easily checked. In
order to fix the Coulomb gauge, we need to add the terms in
Eq.~(\ref{gauge}) and choose the $\alpha_j$'s appropriately. We then have 
\begin{equation}\nonumber
 {\Lambda^{\mu}}_{\nu} A^{\nu} (x') = \int \frac{d\,\mathbf{k}'}{2k_0}\,
 \sum_j
 \frac{\tilde\epsilon_j^{\mu} \hat{a}_j (\Lambda^{-1} k') e^{ik'x'}}{
 \sqrt{(2\pi)^3 2k_0'}} + \mathrm{H.c.},   
\end{equation}
where
\begin{equation}\label{gaugeem}
 \tilde\epsilon_j^{\mu} = {\Lambda^{\mu}}_{\nu} \epsilon_j^{\nu} +
 i\alpha_j k^{\mu}\; . 
\end{equation}
We have to choose $\alpha_j$ such that $\mathbf{k}' \cdot
\tilde{\boldsymbol{\epsilon}}_j =0$. 

We can again extract the annihilation operator of this field, using
the time-independent inner product
\begin{equation}
 \hat{a}'_j (k') = i\int d^3 x'\,
 f^*_{k',j,\mu} (x') \, \overset{\leftrightarrow}{\partial}_{0'}\,
 {\Lambda^{\mu}}_{\nu} A^{\nu} (x')\; ,
\end{equation}
with $f_{k,j,\mu} (x) = \epsilon_{j,\mu}(k)\, [(2\pi)^3
  k_0]^{\frac{1}{2}} \, e^{ikx}$. This leads to the following
  Bogoliubov transformation for polarized light: 
\begin{equation}
 \hat{a}'_j (\Lambda k) =  \sum_l \epsilon^*_{j,\mu} \cdot
 \tilde\epsilon_l^{\mu}\; \hat{a}_l (k)\, e^{-i\Lambda k\ell}\; .
\end{equation}

We will now evaluate $\tilde\epsilon_j^{\mu}$. A Lorentz
transformation $\Lambda$ can be written as a combination of a pure
boost $L$ and two spatial rotations ${\mathcal{R}}_1$ and
${\mathcal{R}}_2$ such that $\Lambda = {\mathcal{R}}_2 L
{\mathcal{R}}_1$.  Note that ${\mathcal{R}}_1$ and ${\mathcal{R}}_2$
are $4\times 4$ matrices of the from $1\oplus R$, with $R$ a $3\times
3$ rotation matrix and 1 the one-dimensional unit matrix. Using
Eq.~(\ref{gaugeem}) with $\Lambda=L$, we find after some algebra that
the polarization of the electromagnetic field in the Coulomb gauge is
{\em not affected by pure Lorentz boosts}. Consequently, with
${\mathcal{R}} \equiv {\mathcal{R}}_2 {\mathcal{R}}_1$ we can write
$\tilde\epsilon_j^{\mu} = {\mathcal{R}}_{\nu}^{\mu} \,
\epsilon_j^{\nu} + i\alpha_j\, k^{\mu}$,  or
$\tilde{\boldsymbol{\epsilon}}_j = R\, \boldsymbol{\epsilon}_j^{\nu} +
i\alpha_j\, \mathbf{k}$.  In addition, pure space rotations leave the
field in both the Coulomb and the Lorentz gauge (i.e.,
$\alpha_j=0$). The transformation law for the annihilation operator or
the electromagnetic field then becomes 
\begin{eqnarray}\label{bogem}
 \hat{a}_j'(\Lambda k) &=& \sum_l \epsilon_j\cdot {\mathcal{R}}\,
 \epsilon_l\; \hat{a}_l (k)\, e^{-i\Lambda k\ell} \cr
 &\equiv& \sum_l U_{jl}\; \hat{a}_l (k)\, e^{-i\Lambda k\ell}\; ,
\end{eqnarray}
with $U_{jl}$ the $2\times 2$ unitary matrix associated with the overall
spatial rotation $R$.

\subsection{Spinor fields}

\noindent 
The spinor field $\psi(x)$ obeys the Dirac equation
$(i\gamma^{\mu}\partial_{\mu}-m)\,\psi = 0$, where $m$ is the mass of
the fermion. We use the gamma matrices $\gamma^{\mu}$ in the standard
representation such that $\gamma^0 = \text{diag}(1,1,-1,-1)$. The
plane-wave solutions to the Dirac equation can then be written as
\begin{equation}\label{eq:spinorfield}
 \psi(x) = \int \frac{d\,\mathbf{k}}{2k_0}\, \sum_{j=1,2}
 \frac{u_j(k)\, \hat{b}_j(k)\, e^{ikx}}{\sqrt{(2\pi)^3 2 k_0}} +
 \frac{v_j(k)\, \hat{d}_j^{\dagger}(k)\, e^{-ikx}}{\sqrt{(2\pi)^3 2
 k_0}} . 
\end{equation}
Here, $\hat{b}_j$ and $\hat{d}_j^{\dagger}$ are the annihilation and
creation operator of the fermion and the anti-fermion in spin state
$j\in\{1,2\}$, respectively. The spinors $u_j(k)$ and $v_j(k)$ are
four-dimensional vectors with $\bar{u}_j \equiv (\gamma^0
u_j)^{\dagger} = u_j^{\dagger} \gamma^0$. The
annihilation operator $\hat{b}_j(k)$ is extracted using the
following time-independent inner product: 
\begin{equation}\label{tiip}
 \hat{b}_j(k) = i\int d^3 x\,
 \frac{\bar{u}_j(k) e^{-ikx}}{\sqrt{(2\pi)^3 2 k_0}}\,
 \overset{\leftrightarrow}{\partial_{0}}\, 
 \psi(x)\; .
\end{equation}

Again, we write the spinor field in terms of Bob's coordinates
and substitute this into the time-independent inner product. The
transformed annihilation operator thus becomes 
\begin{eqnarray}
 \hat{b}_j' (\Lambda k) &\propto& \sum_{l}
 \bar{u}_j(\Lambda k)\, u_{l}(k)\; \hat{b}_{l}(k)\,
 e^{-i\Lambda k\ell} \cr &\equiv& \sum_{l} D_{jl}\;
 \hat{b}_{l}(k)\, e^{-i\Lambda k\ell}, 
\end{eqnarray}
where $D_{jl}$ is a $2\times 2$ unitary matrix up
to a non-unit determinant. This apparent loss of unitarity ($\det
D\neq 1$) is again due to the fact that the spatial integral in
Eq.~(\ref{tiip}) is not Lorentz invariant (three-volumes are not
preserved), and we have to normalize $D$ such that the determinant
becomes one. Using $\det D = \frac{1}{2} [1- k_{\nu} {\Lambda^{\nu}}_{\mu}
  k^{\mu}]$, the Bogoliubov transformation for the spinor annihilation
operator becomes 
\begin{equation}
 \hat{b}_j'(\Lambda k) = \sum_{l} \frac{2\,
  \bar{u}_j(\Lambda k)\, u_{l} (k)}{1-
   k_{\nu}{\Lambda^{\nu}}_{\mu} k^{\mu}}\; \hat{b}_{l}(k)\,
  e^{-i\Lambda k\ell} \; .
\end{equation}
The Bogoliubov transformation for $\hat{d}_j'(k')$ can be
derived along the same lines. Furthermore, it is easily verified that
the fermionic anti-commutation relations are unaltered by Lorentz
transformations.  

The spinors obey the orthogonality relations $\bar{u}_j(k)\,
u_{l}(k) = - \bar{v}_j(k)\, v_{l}(k) =
\delta_{jl}$, and 
\begin{eqnarray}
 u_1(k) &=& \sqrt{\frac{E+m}{2m}}
 \begin{pmatrix}
  1 & 0 & \frac{k_z}{E+m} & \frac{k_x + i k_y}{E+m}
 \end{pmatrix}^T , \cr
 u_2(k) &=& \sqrt{\frac{E+m}{2m}}
 \begin{pmatrix}
  0 & 1 & \frac{k_x - i k_y}{E+m} & \frac{-k_z}{E+m}
 \end{pmatrix}^T .
\end{eqnarray}
Here $E = \sqrt{\mathbf{k}^2 + m^2}$. In this representation it is
straightforward to calculate the matrix elements $\bar{u}_j
(\Lambda k)\,  u_{l}(k)$.

\section{Bosonic and fermionic multi-particle interferometry}\label{sec:int}

\noindent
As was argued in the introduction, two identical single-particle plane
waves incident on a beam splitter will result in bunching (bosons) or
anti-bunching (fermions) in particle detectors in the outgoing modes
of the beam splitter. In the boosted reference frame, these waves will
in general no longer be identical, and as a result one would expect
deviations in the statistics of the detectors. The resolution of this
paradox, both for photons and electrons, will naturally lead to a
unified description of multi-particle boson and fermion interferometry.

\subsection{Twin-photon paradox}

\noindent
First, consider the twin-photon paradox. The transformation law in
Eq.~(\ref{bogem}) indicates that there is no polarization rotation
associated with pure boosts. Therefore, let the boosted annihilation
operator be given by $\hat{a}' (\Lambda k) = \hat{a} (k)$, where we
have chosen $\ell=0$. The boosted single-photon waves can then be
written as $\mbox{$\hat{a}^{\dagger}$} (\Lambda k_1)
\mbox{$\hat{a}^{\dagger}$} (\Lambda k_2)|0\rangle$. It is clear that 
the frequency components of $\Lambda k_1$ and  $\Lambda k_2$ will
generally be different. However, in the boosted frame, the beam
splitter action will also change: We can write the interaction
Hamiltonian of the beam splitter at rest with incoming modes $k_1$ and
$k_2$ as the bilinear form 
\begin{equation}
 H = \frac{\pi i}{2} \left[ \hat{a}^{\dagger}(k_1)\, \hat{a}(k_2) -
 \hat{a}(k_1)\, \hat{a}^{\dagger}(k_2) \right].
\end{equation} 
In the boosted frame, this Hamiltonian will become 
\begin{equation}\nonumber
 H' = \frac{\pi i}{2} \left[ \mbox{$\hat{a}'$}^{\dagger} (\Lambda
 k_1)\, \hat{a}' (\Lambda k_2) - \hat{a}'(\Lambda k_1)\,
 \mbox{$\hat{a}'$}^{\dagger} (\Lambda k_2) \right].
\end{equation} 
The linear Bogoliubov transformation corresponding to the beam 
splitter action in the boosted frame is then given by $\exp(iH')\,
\hat{a}' (\Lambda k_i) \exp(-iH')$. Using the Baker-Campbell-Hausdorff
relation 
\begin{equation}\nonumber
 e^{\lambda A} B e^{-\lambda A} = B + \lambda [A,B] + \frac{\lambda^2}{2}
 \left[ A,[A,B] \right] + \ldots, 
\end{equation}
we find that the boosted beam splitter transformation becomes  
\begin{eqnarray}
 \hat{a} (\Lambda k_1) &\rightarrow& \frac{1}{\sqrt{2}} \left( \hat{a}'
 (\Lambda k_1) - \hat{a}' (\Lambda k_2) \right) \cr
 \hat{a} (\Lambda k_2) &\rightarrow& \frac{1}{\sqrt{2}} \left( \hat{a}'
 (\Lambda k_1) + \hat{a}' (\Lambda k_2) \right) \; .
\end{eqnarray}
In other words, the boosted beam splitter induces an interaction
between different frequencies of the incoming field. The beam splitter
thus defines a preferred frame of reference in which identical photons
exhibit the Hong-Ou-Mandel effect. This resolves the twin-photon
paradox. 

\subsection{Twin-electron paradox}

\noindent
The fermionic beam splitter is in many ways the dual of the bosonic
case: When two identical fermions enter the two input ports of a
fermionic beam splitter, the exclusion principle dictates that they
will never leave the same output port. In the boosted frame, however,
both the wavelength and the spins of the fermions change. Naively,
this again results in a way of distinguishing the particles, with a
change in detector statistics as a result. 

The resolution of this paradox is similar to that of the twin-photon
paradox: we need to transform the interaction Hamiltonian of the
fermionic beam splitter. In addition, we need to show that the spin
rotation does not actually render the particles distinguishable. Due
to the fermionic anti-commutation relation $\{ \hat{b}_{j} (k_1), 
\hat{b}_{l} (k_2) \} = 0$, a 50:50 electron beam splitter
transforms the input state $\hat{b}^{\dagger}_{j} (k_1)\,
\hat{b}^{\dagger}_{j} (k_2) |0\rangle$ into itself. In the
boosted frame, the transformation becomes
\begin{eqnarray}\nonumber
  \mbox{$\hat{b}'$}^{\dagger}_j (\Lambda k_1)\,
  \mbox{$\hat{b}'$}^{\dagger}_j (\Lambda k_2) &\rightarrow& 
  \sum_{l} D_{jl}^2\,
  \mbox{$\hat{b}'$}^{\dagger}_{l} (\Lambda k_1)\,
  \mbox{$\hat{b}'$}^{\dagger}_{l} (\Lambda k_2) \cr && + 
  \frac{1}{2}\sum_{l\neq m} D_{jl} D_{jm} 
  \mbox{$\hat{b}'$}^{\dagger}_{l} 
  (\Lambda k_1)\,  \mbox{$\hat{b}'$}^{\dagger}_{m} (\Lambda k_1)
  \cr && 
  - \frac{1}{2}\sum_{l\neq m} D_{jl} D_{jm} 
  \mbox{$\hat{b}'$}^{\dagger}_{l} (\Lambda k_2)\,
  \mbox{$\hat{b}'$}^{\dagger}_{m} (\Lambda k_2). 
\end{eqnarray}
The fermionic anti-commutation relations render the last two sums
zero. In the boosted frame, the state is then
\begin{equation} 
 |\psi\rangle = \sum_{l} D_{jl}^2\,
 \mbox{$\hat{b}'$}^{\dagger}_{l} (\Lambda k_1)\,
 \mbox{$\hat{b}'$}^{\dagger}_{l} (\Lambda k_2) |0\rangle\; .
\end{equation} 
In other words, the fermions always occupy different spatial modes,
and they have the same (boosted) spin. Note that it is not the details
of $D$ that resolve the paradox, but the canonical commutation
relations. This mechanism is indicative of other massive quantum
fields as well: the paradox for massive bosons is resolved by   
invoking the bosonic commutation relations.

\subsection{Non-inertial movement and quantum interferometry}

\noindent
It should be stressed that the resolution of the twin-particle
paradox, as sketched above, is valid only for inertial observers. In
other words, when Alice and Bob are in non-inertial motion with
respect to each other, they generally cannot agree upon a shared
vacuum state. As a consequence, Alice and Bob no longer agree on the
number of particles that are involved in the experiment. In physical
terms, when Alice prepares the experiment to demonstrate the
Hong-Ou-Mandel effect, Bob will see a beam splitter that emits thermal
radiation. The energy of this radiation will be supplied by the
mechanism that drives the beam splitter away from inertial motion 
(according to Bob).

\subsubsection{Bosons}

Nevertheless, the transformation properties of the field operators
allow a Hamiltonian formulation of multi-particle quantum interferometry for
non-inertial observers. We will sketch a proof that the Hamiltonian of
a linear interferometer (including squeezing) is properly transformed by 
substituting the Bogoliubov transformation of the creation and
annihilation operators. We use the fact that the Hamiltonian of a free
(scalar) field $\phi$ can be written as 
\begin{equation}\nonumber
 H = \int d^3 x\, T_{00} = \frac{1}{2} (i \partial_0 \phi,\phi), 
\end{equation}
with $T_{00}$ the Hamiltonian density component of the stress tensor
$T$, and
\begin{equation}\nonumber
  \phi = \int \frac{d\,\mathbf{k}}{2k_0}\, (f_k \hat{a}_k + f_k^*
  \hat{a}^{\dagger}_k) = \int \frac{d\,\mathbf{k}}{2k_0}\, (g_k
  \hat{b}_k + g_k^* \hat{b}^{\dagger}_k).   
\end{equation} 
Here $f_k$ and $g_k$ are orthonormal mode functions corresponding to
ladder operators $\hat{a}_k \equiv \hat{a}(k)$ and $\hat{b}_k \equiv
\hat{b}(k)$, respectively, and $(f_j,f_k) = - (f^*_j,f^*_k) =
\delta_{jk}$ and $(f^*_j,f_k)=0$. Furthermore, we can write the
annihilation operator in terms of the inner products of the mode
functions and the transformed creation and annihilation operators:
\begin{equation}
  \hat{a}_k = \int \frac{d\,\mathbf{k}}{2k_0}\, [(g_j,f_k) \hat{b}_j -
 (f^*_k,g_j) \hat{b}_j^{\dagger}].
\end{equation}
Substituting this transformation rule into $H(\hat{a},
\hat{a}^{\dagger})$ and using the completeness relation
\begin{equation}\nonumber
 (\phi_1,\phi_2) = \int dk [(\phi_1,f_k)(f_k,\phi_2) -
    (\phi_1,f_k^*)(f_k^*,\phi_2)] 
\end{equation}
yields the transformed Hamiltonian $H(\hat{b},\hat{b}^{\dagger})$. 

In order to find the relativistic extension to linear multi-particle
quantum interferometry, we generalize this result to any $N$-mode
bilinear form of the interferometer's interaction Hamiltonian 
\begin{equation}
 H = \frac{1}{2} \sum_{j,k=1}^N \left( \hat{a}^{\dagger}_j
 A_{jk}\hat{a}^{\dagger}_k + 2 \hat{a}^{\dagger}_j B_{jk}\hat{a}_k +
 \hat{a}_j A_{jk} \hat{a}_k \right) , 
\end{equation}
where $A$ and $B$ are symmetric $N\times N$ Hermitian matrices. They
give a complete description of $N$-mode multi-particle interferometry,
including 
multi-mode squeezing (particle creation). Such Hamiltonians yield
linear Bogoliubov transformations of the annihilation operators. It
follows from Eq.~(\ref{tii}) that any {\em arbitrary coordinate
  transformation} of the quantum fields (including non-inertial
transformations) also gives a Bogoliubov transformation of the
annihilation  and creation operators, and as a result preserves the
structure of this Hamiltonian. So far, we have derived the
relativistic generalization of multi-particle interferometry for scalar
fields. Optical interferometry requires that we take into account the
extra gauge freedom of the electromagnetic field. General coordinate
transformations will induce a nontrivial polarization rotation. It is
easy to see, however, that the resulting Bogoliubov transformations
allow for a similar relativistic extension.

There is, however, one subtlety: When non-inertial observers are
involved, the parties in question typically do not have access to all
the modes in the Hamiltonian. Some modes will be causally separated by
(effective) event horizons. Consequently, the field amplitudes of
these modes have to be traced out, and the observers will find that
the output of the interferometer is in a mixed state.
  
\subsubsection{Fermions}

We can generalize the {\em fermionic} beam splitter to arbitrary
$N$-mode fermionic interferometers in a similar way. To 
include the spin-$\frac{1}{2}$ degree of freedom, we assume $M$
spatial modes, such that $N=2M$. Furthermore, each mode can be
occupied by both electrons and anti-electrons (denoted by $b$ and $d$,
respectively). From the expansion of the spinor field in
Eq.~(\ref{eq:spinorfield}) we see that general Bogoliubov
transformations have the form 
\begin{eqnarray}
 \hat{b}_j &\rightarrow& \sum_k U_{jk} \hat{b}_k + V_{jk}
 \hat{d}_k^{\dagger} \; ,\cr
 \hat{d}_j^{\dagger} &\rightarrow& \sum_k W_{jk} \hat{d}_k^{\dagger} +
 Z_{jk} \hat{b}_k\; ,
\end{eqnarray}
with $U$, $V$, $W$, and $Z$ determined by the coordinate transformation.
The bilinear interaction Hamiltonian of the free spinor field must
be closed under these transformations. To this end, we define the
$2N$-tuple of creation and annihilation operators $\vec{s} \equiv
(\hat{b}_1,\ldots,\hat{b}_N,\hat{d}_1^{\dagger} ,  \ldots,
\hat{d}_N^{\dagger} )$. The Hamiltonian for multi-particle fermion
interferometry is then given by 
\begin{equation}
 H = \frac{1}{2} \sum_{j,k=1}^{2N} \left( \hat{s}^{\dagger}_j
 {\mathcal{A}}_{jk}\hat{s}^{\dagger}_k + 2 \hat{s}^{\dagger}_j
 {\mathcal{B}}_{jk}\hat{s}_k + \hat{s}_j {\mathcal{A}}_{jk} \hat{s}_k
 \right) ,  
\end{equation}
with
\begin{equation}
 {\mathcal{A}} \equiv 
 \begin{pmatrix}
  ~A_1 & C \cr
  -C & A_2 
 \end{pmatrix} \quad\text{and}\quad
 {\mathcal{B}} \equiv
 \begin{pmatrix}
  ~B_1 & D \cr
  -D & B_2 
 \end{pmatrix} ,
\end{equation}
where $A_{1,2}$, $B_{1,2}$, $C$ and $D$ are anti-symmetric $N\times
N$ Hermitian matrices. This constitutes a unified theory of
relativistic multi-particle fermionic interferometry.

\section{Conclusions}\label{sec:con}

\noindent
In conclusion, we have derived the explicit form of the annihilation
operator of some common quantum fields under relativistic
transformations. Our results are completely general and offer a
straightforward and fundamental way to calculate relativistic effects
in bosonic and fermionic interferometers, thus establishing a unified
theory of relativistic multi-particle quantum interferometry and
linear optical quantum computing. With this theory, we can describe
the behavior of  multi-particle wave packets rather than momentum
eigenstates in a direct manner. Most notably, our technique can be
applied to arbitrary coordinate transformations, and can be used to
study entanglement and relativistic quantum information theory in
arbitrary coordinate systems.


\section*{Acknowledgments}

\noindent
P.K.\ is supported by the European Union RAMBOQ project, and S.L.B.\
currently holds a Royal Society-Wolfson research merit award.

\bibliography{rf}

\end{document}